# Organizational economic sustainability via process optimization and human capital: a Soft Systems Methodology (SSM) approach

**Wadim Strielkowski [1,2] *, Evgeny Kuzmin [3], Arina Suvorova [4, 5], Natalya Nikitina [6] and Olga Gorlova [7]**

1 Department Agricultural and Resource Economics, University of California, Berkeley, 303 Giannini Hall, CA 94720, United States; strielkowski@berkeley.edu
2 Department of Trade and Finance, Faculty of Economics and Management, Czech University of Life Sciences Prague, Kamýcká 129, Prague 6, 165 00 Prague, Czech Republic; strielkowski@pef.czu.cz
3 Department of Regional Industrial Policy and Economic Security, Institute of Economics of the Ural Branch of the Russian Academy of Sciences, Moskovskaya str. 29, 620014 Ekaterinburg, Russia; kuzmin.ea@uiec.ru
4 Laboratory of Economic Genetics of Regions, Institute of Economics of the Ural Branch of the Russian Academy of Sciences, Moskovskaya str. 29, 620014 Ekaterinburg, Russia; suvorova.av@uiec.ru
5 Graduate School of Economics and Management, Ural Federal University, Mira str. 19, 620002 Ekaterinburg, Russia, suvorova.av@uiec.ru
6 Department of Modern Pedagogy, Lifelong Education and Personal Tracks, Russian State Social University, Wilhelm Pieck str. 4/1, 129226 Moscow, Russia; nikitina@ymservices.ru
7 Department of Advertising and Public Relations in the Media Industry, Moscow Polytechnic University, Bolshaya Semyonovskaya str. 38, 107023 Moscow, Russia; business007@bk.ru

* Correspondence: strielkowski@berkeley.edu

**Abstract:** This review paper focuses on enhancing organizational economic sustainability through process optimization and human capital effective management utilizing the soft systems methodology (SSM) approach which offers a holistic approach for understanding complex real-world challenges. By emphasizing systems thinking and engaging diverse stakeholders in problem-solving, SSM provides a comprehensive understanding of the problem's context and potential solutions. The approach guides a systematic process of inquiry that leads to feasible and desirable changes in tackling complex problems effectively.
Our paper employs the bibliometric analysis based on the sample of 5171 research articles, proceedings papers, and book chapters indexed in Web of Science (WoS) database. We carry out the network cluster analysis using the text data and the bibliometric data with the help of VOSViewer software. Our results confirm that as the real-world situations are becoming more complex and the new challenges such as the global warming and climate change are threatening many economic and social processes, SSM approach is currently getting back at the forefront of academic research related to such topics as organizational management and sustainable human capital efficiency.

## 1. Introduction

In the post-COVID era, the need for increasing environmental awareness and adapting sustainable practices leads to the renaissance of the socio-economic approaches such as human capital management. Soft Systems Methodology (SSM), a method that was developed in the 1960s and represented the structured way of addressing complex real-world challenges comes, is finding a new role in this process.

SSM sees the issue as not some isolated entities but as being interconnected with other elements within their respective environment. It seeks to identify these relationships and interactions to gain a comprehensive understanding of the problem at hand [1, 2]. One key feature of SSM is its emphasis on learning from diverse stakeholders. Unlike traditional approaches that rely solely on expert knowledge or quantitative data, SSM recognizes the value of different perspectives and experiences in problem-solving. By engaging stakeholders from various backgrounds, such as managers, employees, customers, or community members, SSM aims to build a shared understanding of the problem's context and potential solutions [3, 4]. In order to lead this process, SSM uses a set of tools and techniques known as the "seven-stage model" [5] (see Figure 1).

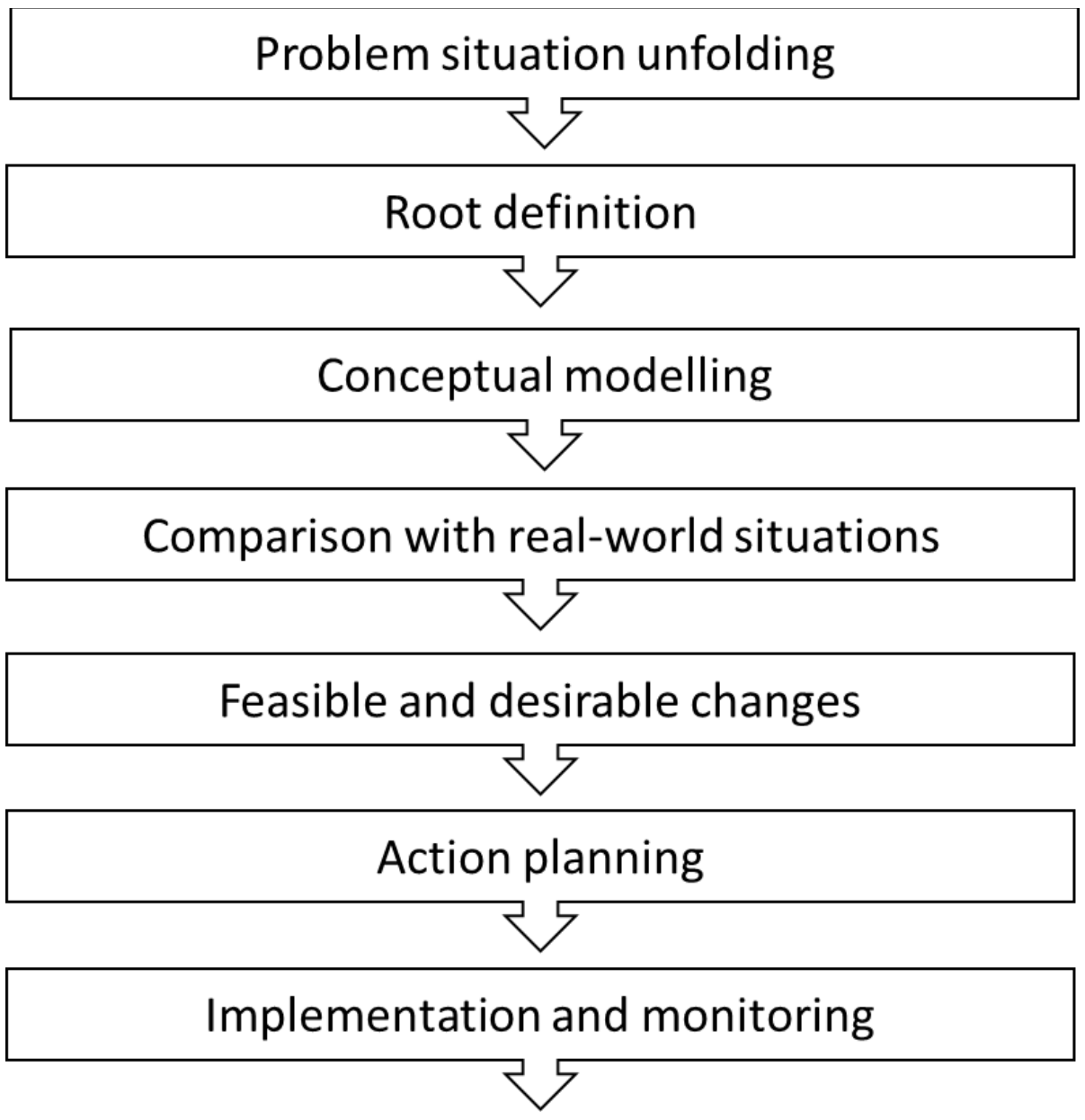

**Fig. 1.** SSM seven-stage model. Source: Own results

As outlined in Figure 1, these seven stages of the SSM model provide a systematic framework for exploring complex situations:

1) *Problem situation unfolding*: identifying and defining the perceived problem situation from different viewpoints;
2) *Root definition*: stakeholders have to work together to develop an agreed-upon root definition that captures the essential elements of the problem situation;
3) *Conceptual modeling*: constructing conceptual models representing different perspectives within the system under study;
4) *Comparison with real-world situations*: conceptual models are then compared with real-world situations to identify gaps or discrepancies;
5) *Feasible and desirable changes*: stakeholders collaborate to generate potential changes or improvements that are both feasible and desirable;
6) *Action planning*: developing detailed plans for implementing the identified changes, considering practical constraints, and resources;
7) *Implementation and monitoring*: putting the action plans into practice, monitoring their effectiveness, and making adjustments as needed.

By following this seven-stage model, SSM facilitates a structured process of exploration, learning, and decision-making. It helps stakeholders to navigate complex problems by breaking them down into manageable parts while still considering the interrelationships between these parts. It can also help with effectively managing human capital in organizations and institutions [6, 7].

In the same time, here is the human capital which referrs to the knowledge, skills, and abilities of an organization's workforce, is a critical asset that can significantly impact its overall sustainability [8, 9]. The concept of enhancing sustainability in human capital management involves optimizing processes and strategies to ensure the long-term well-being of employees while maximizing organizational effectiveness [10]. This approach recognizes that employees are not only valuable resources but also stakeholders with legitimate expectations for fair treatment and opportunities for growth. By adopting a SSM approach, organizations can systematically analyze their current human capital management practices and identify areas for improvement [11]. This is due to the fact that SSM is a problem-solving technique that emphasizes understanding complex systems from multiple perspectives to facilitate effective decision-making. It enables organizations to consider various stakeholder viewpoints and incorporatetheir interests into sustainable human capital management strategies [12, 13]. One of the key benefits of enhancing sustainability in human capital management is improved employee engagement and satisfaction. When employees feel valued and supported by their organization, they are more likely to be motivated, productive, and committedto its success [14, 15]. Sustainable practices such as providing work-life balance initiatives, fostering a diverse and inclusive work environment, offering career development opportunities, and promoting health and well-being can contribute significantly to employee satisfaction [16, 17].

Hence, enhancing sustainability in human capital management is not only beneficial for employees but also critical for organizational success. By adopting a SSM approach, organizations can identify areas for improvement and develop sustainable practices that enhance employee engagement, attract top talent, and build long-term resilience [18-20].

This main scientific value-added of this paper is the comprehensive literature review and bibliometric network cluster analysis of the organizational sustainability through process optimization and human capital effective management utilizing the SSM approach. This approach surely has some limitations since the bibliometric analysis is based on Google Trends tools as well as the VOSviewer network analysis of research articles, proceedings papers, and book chapters indexed in Web of Science (WoS) database (while other databases such as Scopus or PubMed were omitted for the sake of saving time and resources). Nevertheless, our approach still yields valuable and informative results that can be used by the researchers and practitioners alike.

## 2. Data and methodology

In order to demonstate that SSM constitutes the comprehensive framework for optimizing HCM processes in organizations and enterprises with a focus on sustainable development, we turn to the bibliometric analysis of the research publications on these topics. We have selected the Web of Science (WoS) database as one of the most prestigious and complete abstract and citation databases that features a vast amount of research on this and related topics. Table 1 describes the distribution of document types (articles, proceedings papers, review articles, and others) used for our analysis.

**Table 1.** Summary of data and data selection algorithm

| Category | Specific criteria |
|---|---|
| Reference and citation database | Web of Science |
| Citation indices | SCI-Expanded, SSCI |
| Time period | 1984-2023 |
| Language | "English" |
| Keywords | "soft systems methodology" |
| Document types: | |
| Articles | 3540 |
| Proceeding papers | 1464 |
| Review articles | 278 |
| Others | 111 |
| Sample size | N = 5171 |

Source: Own results

Moreover, Figure 2 that is shown below provides the description of our algorithm for the data selection, retrieval, processing, as well as the network analysis used in this paper.

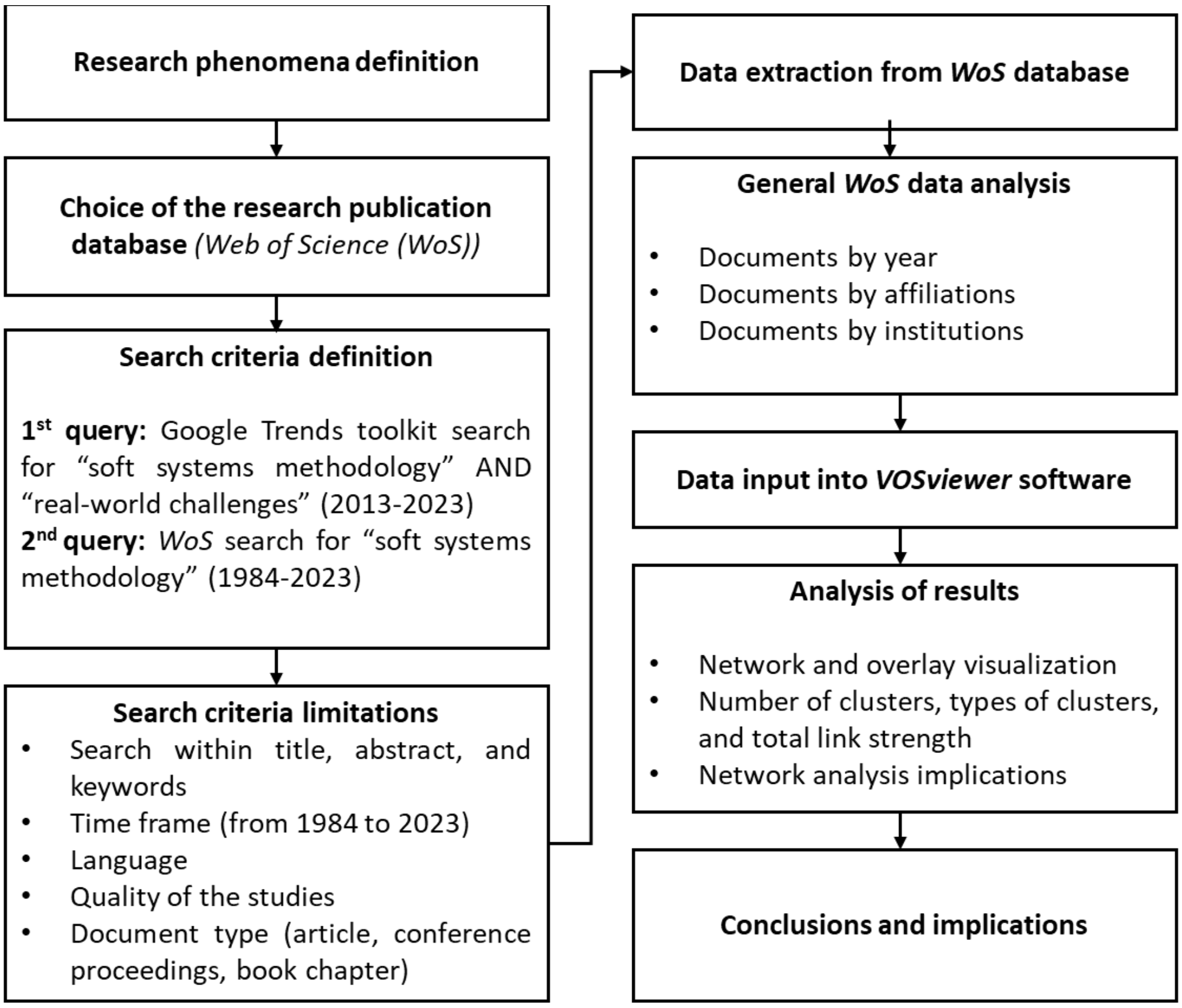

**Fig. 2.** Diagram of the data selection and network analysis algorithm. Source: Own results

Before our data collection from WoS, we run an online analysis using the Google Trends toolkit. Google Trends show the popularity of specific terms in various parts of the world measured by the online searches [21].

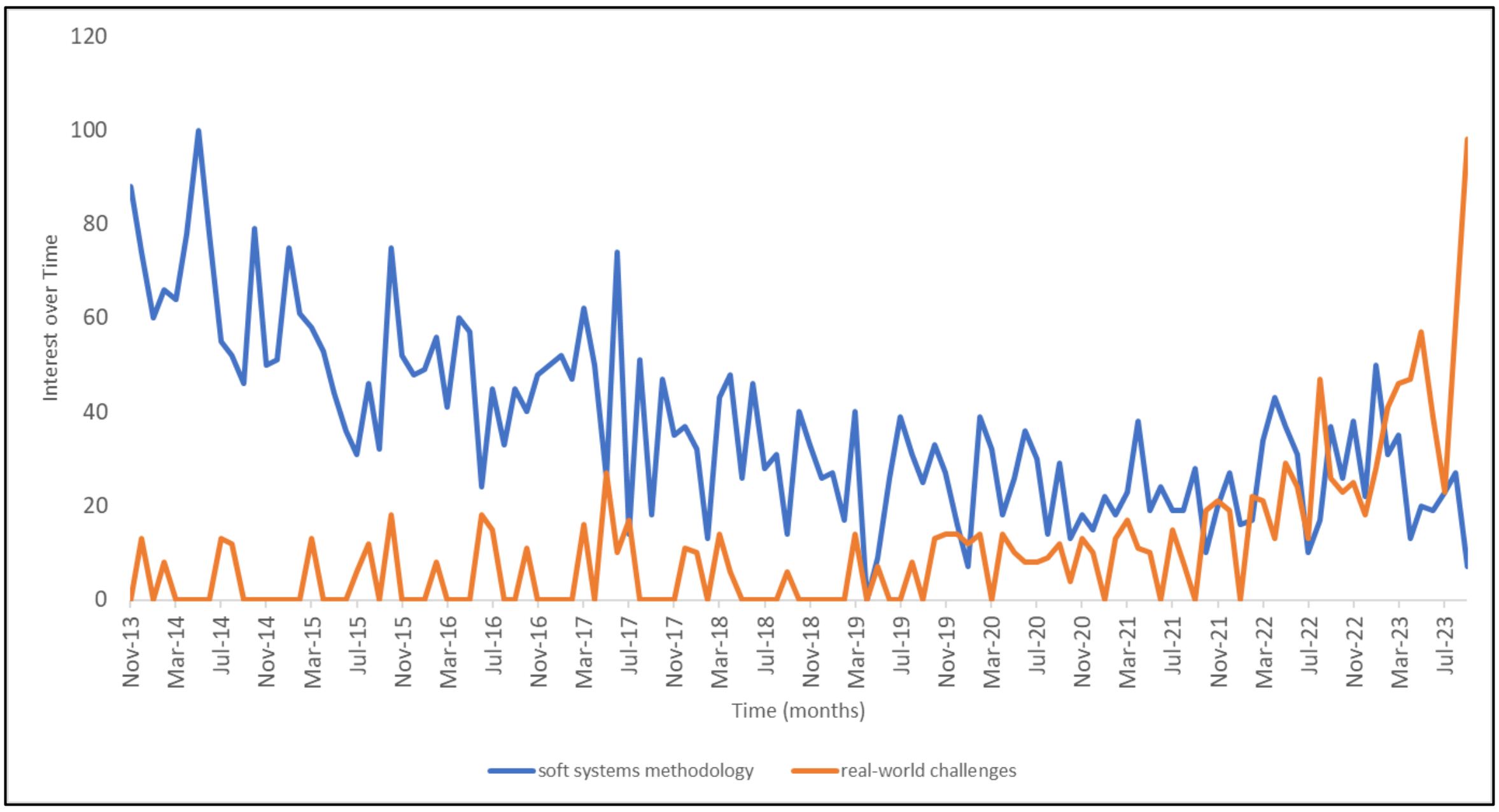

**Fig. 3.** Google Trends of the publications on "soft systems methodology" and "real-world challenges" (2013-2023). Source: Own results based on Google Trends [21]

Figure 3 above reveals the dynamics of the frequency of worldwide search requests using the search items "soft systems methodology" and "real-world challenge" (Figure 3). One can see that the search frequency for both terms was quite different before 2019-2021 with the prevailance of interest in SSM. However, after 2021 the Interest over Time (search interest relative to the highest point for the given region and time) for “real-world challenge” has rapidly increased.

In addition, in the empirical part of our review article that follows, we perform the statistical analyses on the publications indexed in WoS featuring such information as countries, authors, abtracts, and keywords with the help of assessing the co-occurrences and keywords’ cluster analyses.

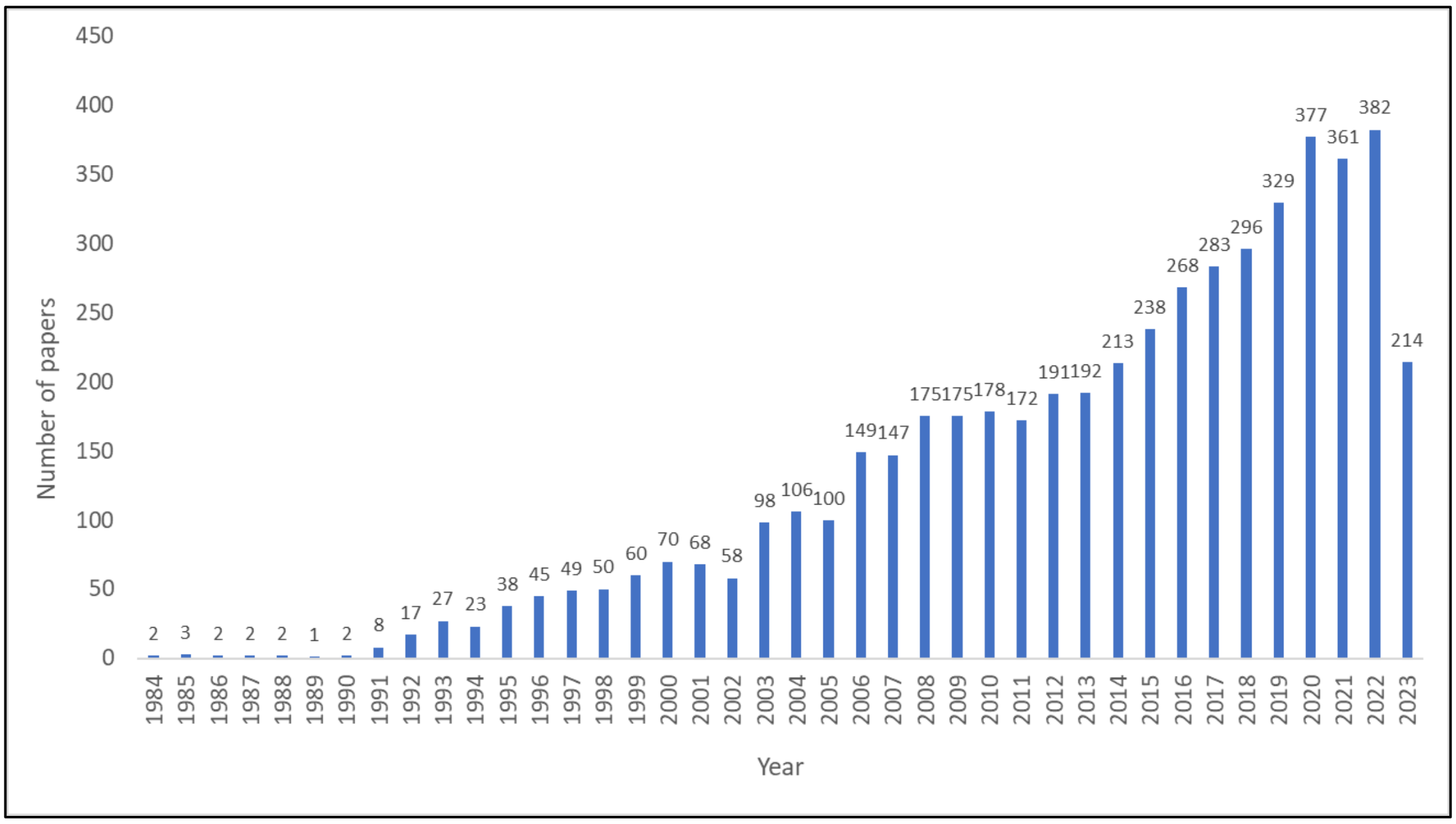

**Fig. 4.** Trend of the publications on soft systems methodology over time (1984-2022). Source: Own results

We made a search in the Web of Science (WoS) database using the terms “soft systems methodology” which left us with a total number of 5171 results from Web of Science Core Collection (3540 articles and 1464 proceeding papers among them). Figure 4 depicted above reports the major trend of the publications on soft systems methodology over time using the sample of 5171 publications from WoS database from 1984 until 2023.

Looking at the trends depicted in Figure 4, it is clear that the main surge in the publications on soft systems methodology occurred around 2003-2005 with the numbers still proportionally increasing (the slump that can be observed in 2023 is due to the fact that the actual number of publications indexed in WoS is not yet known and is still being counted in WoS).

## 3. Main resuts: network cluster analysis

In this section, we present the results of the empirical model based on the bibliometric network cluster analyzis that employs the VOSviewer software which has recently gained popularity for analysing bibliometric data [22, 23]. The results of our analysis are presented in the form of visual network maps that allow us to determine the key trends and patterns.

Bibliometric analysis, a quantitative method for studying patterns within academic literature, offers valuable insights into the evolving landscape of research. In this study, a bibliometric network cluster analysis was conducted on a sample of 5171 publications indexed in the Web of Science database from 1984 to 2023. By employing keywords and phrases associated with "soft systems methodology" (SSM), the analysis revealed four distinct clusters, shedding light on the multifaceted aspects of SSM research

Figure 5 presents the visualization of the network cluster analysis with a map based on the text data from the sample of 5171 publications indexed in WoS database from and published from 1984 until 2023. Our results of the bibliometric network analysis demonstrate that four main clusters were identified. The analysis of using keywords and phrases in the publications retrieved from WoS revealed that the key terms connected to "soft systems methodology” are most often associated with the following concepts: (i) Soft Systems Methodology (Cluster 1 or red clustering); (ii) Property measurement (Cluster 2 or green clustering); (iii) Efficiency design (Cluster 3 or olive color clustering); and iv) Algorithm optimization (Cluster 4 or blue color clustering).

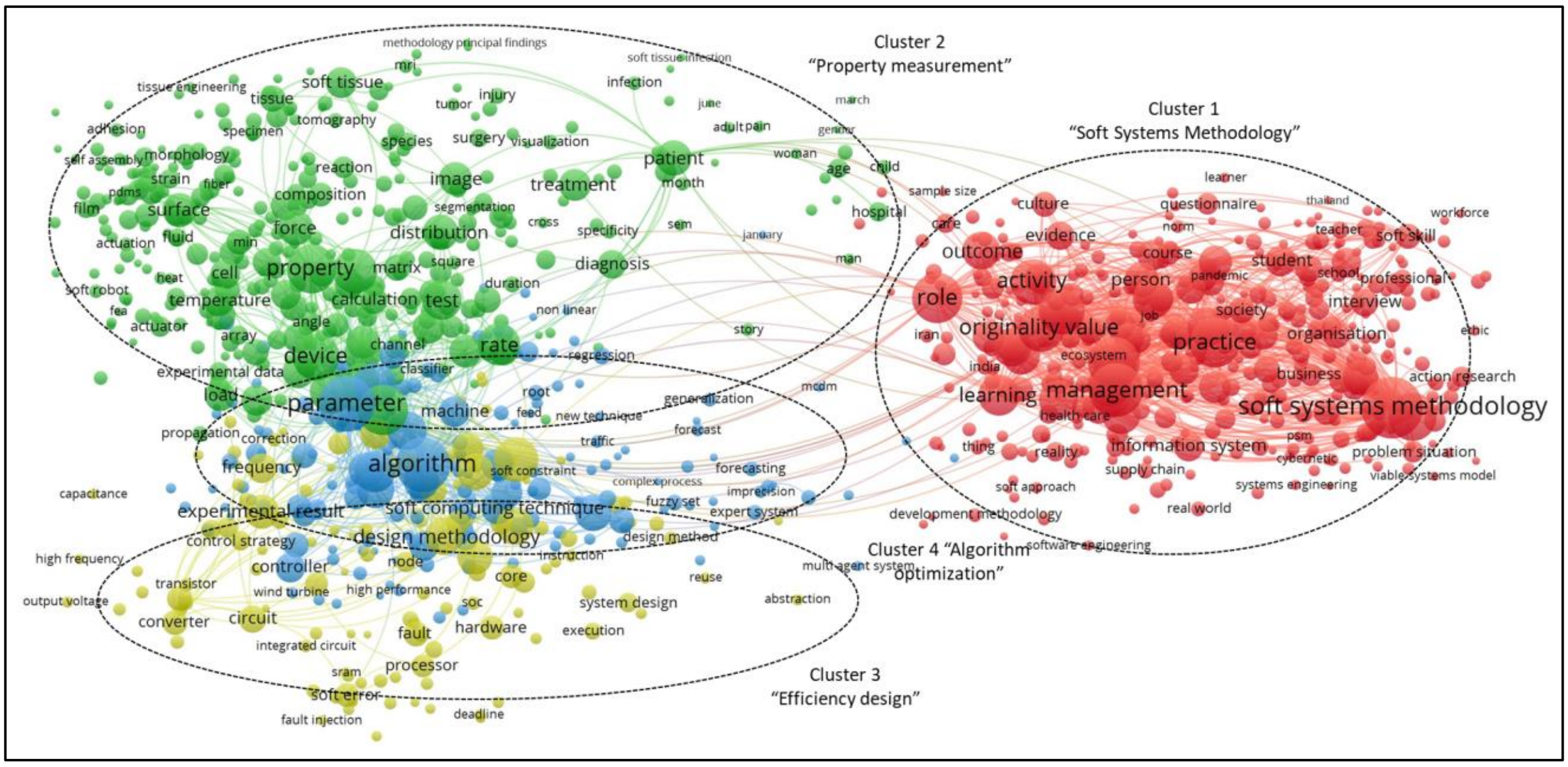

**Figure 5.** The dominant clusters of cross-sector research on SSM retrieved from the sample of 5171 publications indexed in WoS. Source: Own results based on VOSViewer v. 1.6.18 software

The first cluster is centered around the core concept of Soft Systems Methodology itself. Within this cluster, seminal works, theoretical frameworks, and applications of SSM were explored. Researchers delved into the historical development of SSM, examining its roots in system thinking and its evolution as a problem-solving approach. This cluster served as the foundational hub, linking various sub-fields and applications within the broader context of SSM.

The green cluster, focusing on property measurement, reflected a specific area of interest within the realm of SSM. Researchers in this cluster explored methodologies for measuring properties within complex systems. This included the development of novel metrics, indices, and evaluation techniques. The emphasis here was on establishing quantifiable measures to assess the effectiveness of SSM in diverse contexts, fostering a more rigorous and data-driven approach to soft systems research.

The olive-colored cluster, dedicated to efficiency design, delved into the practical applications of SSM in enhancing system efficiency. Scholars within this cluster examined how SSM could be applied to optimize processes, streamline workflows, and improve overall organizational efficiency. Case studies and empirical research played a significant role, demonstrating the real-world impact of SSM methodologies on various industries and sectors

The blue cluster, centered around algorithm optimization, showcased the intersection of SSM with computational techniques. Researchers explored how algorithms could be optimized and integrated into SSM frameworks to enhance decision-making processes. This cluster highlighted the symbiotic relationship between soft methodologies and cutting-edge computational tools, emphasizing the role of technology in advancing the effectiveness of SSM applications

Figure 6 that is shown below the discussion above by offering the density visualization of the network cluster analysis map. It provides the density level of the SSM based on the quantity and depth of research (the thicker the density color, the more research was recorded and documented on this topic or concept) (see Figure 6).

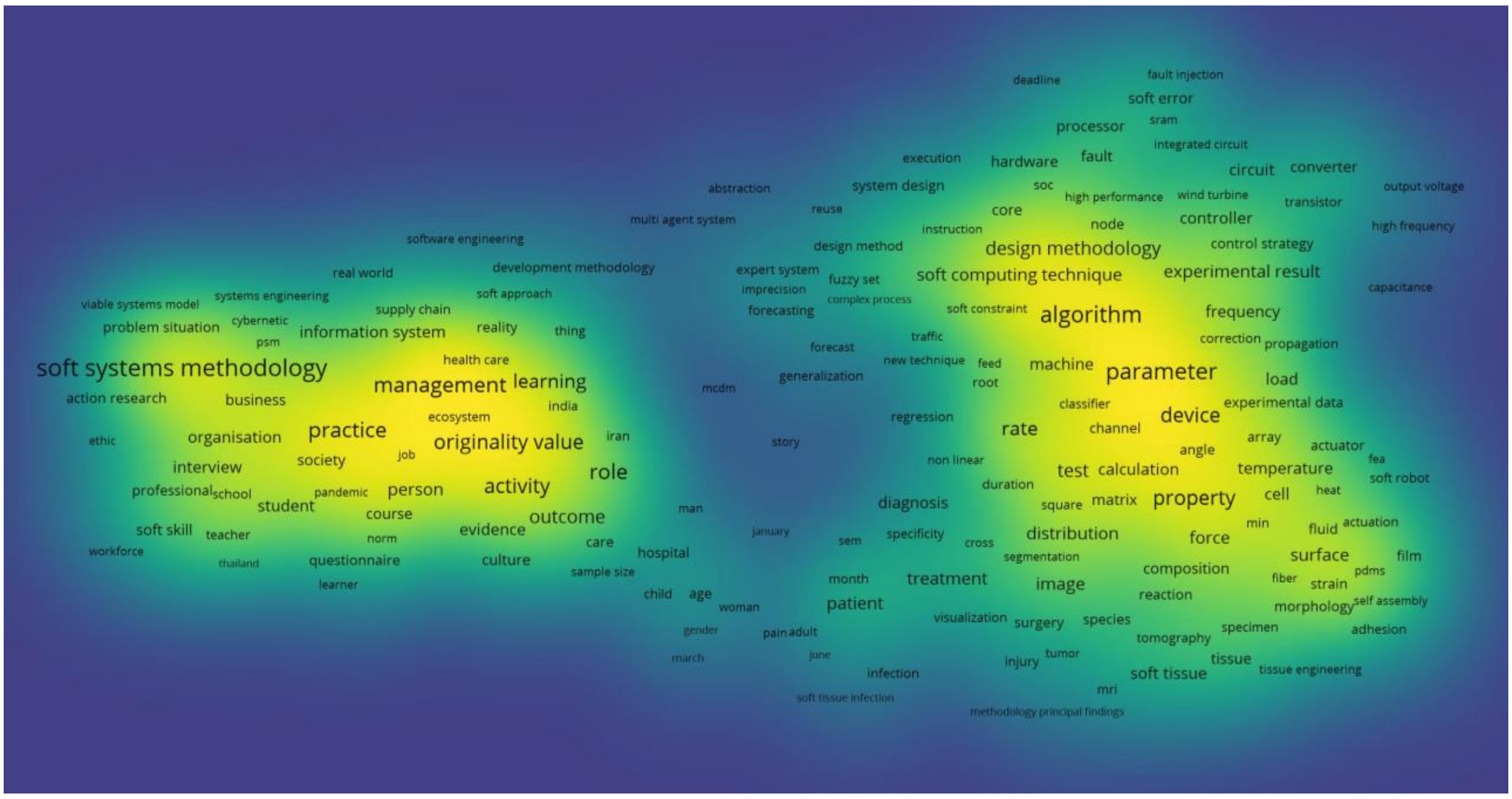

**Figure 6.** Density visualization of the network cluster analysis of the sample of 5171 publications on SSM indexed in WoS. Source: Own results based on VOSViewer v. 1.6.18 software.

Furthermore, Figure 7 reveals the results of the network map based on the bibliographic data (keyword co-occurrences, citation, and bibliographic coupling). From Figure 7 it becomes clear that soft computing and optimization (green clustering), behavior (red clustering), and human capital management (blue clustering) are often mentioned when discussing the provisions of SSM.

The green clustering emerged as a focal point, emphasizing the synergy between SSM and soft computing techniques. Soft computing, encompassing methodologies like neural networks, fuzzy logic, and genetic algorithms, finds resonance with SSM due to its adaptable, human-like problem-solving abilities. Researchers explored the fusion of these computational paradigms with SSM, paving the way for innovative problem-solving approaches. Optimization techniques within this cluster indicated a keen interest in refining SSM applications, ensuring the methodology's efficiency in addressing complex, real-world issues.

The red clustering, centering around behavior, delved into the human dimension of SSM. Recognizing the pivotal role of human behavior in shaping organizational dynamics, researchers explored behavioral aspects within the SSM framework. This cluster delved into psychological and sociological factors influencing decision-making processes within soft systems. Understanding human behavior became integral to enhancing the efficacy of SSM applications, emphasizing the importance of aligning methodologies with human cognitive processes and social contexts.

The blue clustering, focusing on human capital management, underscored the significance of skilled professionals in the successful implementation of SSM. Human capital, referring to the collective knowledge, skills, and expertise of an organization's workforce, emerged as a critical factor in SSM discourse. Researchers examined strategies for harnessing human capital effectively within the SSM framework. This cluster highlighted the role of education, training, and talent management in ensuring the seamless integration of SSM into diverse organizational settings.

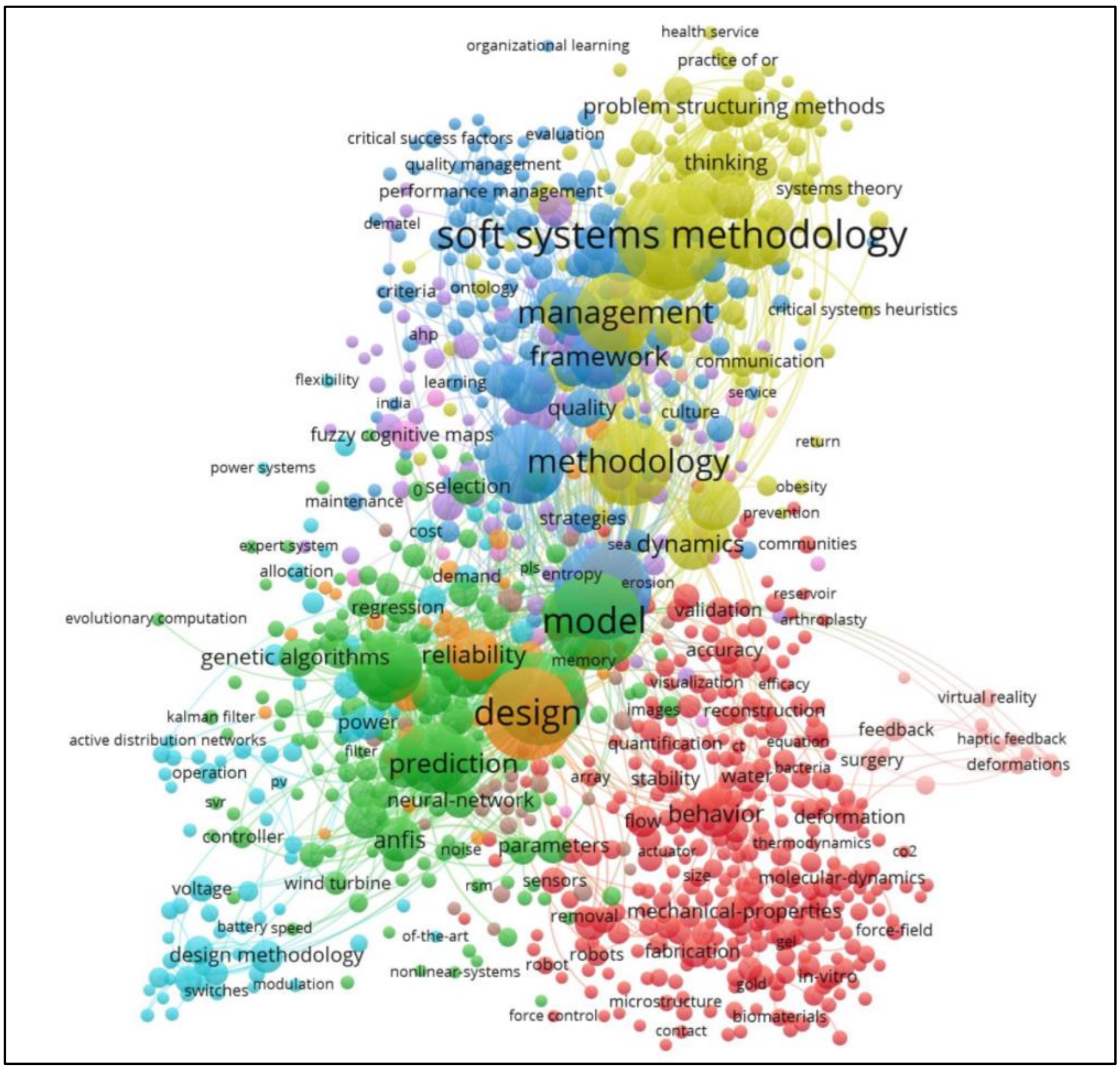

**Figure 7.** Network map based on the bibliographic data of the sample of papers containing the keywords "SSM" retrieved from the sample of 5171 publications indexed in WoS.
Source: Own results based on VOSViewer v. 1.6.18 software.

The results of our bibliometric network cluster analysis provided a comprehensive overview of the diverse research landscape related to soft systems methodology. By identifying these distinct clusters, researchers gained valuable insights into the multifaceted nature of SSM research. From foundational explorations of SSM principles to the practical applications in efficiency design and the integration of computational algorithms, the clusters reflected the interdisciplinary and evolving nature of SSM research. This analysis not only facilitated a deeper understanding of the existing body of knowledge but also paved the way for future research directions, encouraging scholars to explore the synergies between SSM and emerging technologies, measurement methodologies, and efficiency-driven applications in complex systems. Through the use of cluster analyses conducted in this section, it becomes evident that SSM holds substantial importance in the context of the emergence of exceptional scholarly investigations.

## 8. Conclusions and implications

Overall, it appears that Soft Systems Methodology (SSM) offers a promising solution fof the organizational economic sustainability via process optimization and human capital by providing a structured framework for understanding and tackling complex problems. Even though the methodology is more than 60 years old, it still offers a holistic approach for addressing complex real-world challenges by embracing key principles such as understanding the problem situation, defining relevant systems, creating conceptual models, comparing real-world systems with models, identifying desirable changes, formulating feasible actions, facilitating learning and adaptation, as well as promoting collaboration and participation. These principles enable stakeholders to navigate through complexity while fostering creativity, innovation, shared understanding, and sustainable solutions to address complex challenges effectively.

One of the key strengths of SSM lies in its ability to uncover underlying assumptions and mental models that influence problem perception. By engaging stakeholders in rich dialogue, SSM encourages individuals to challenge their preconceived notions about how things should work. This reflective process enables participants to gain new insights into complex situations and opens up possibilities for creative problem-solving

The results stemming from our bibliometric network cluster analysis illuminated the multidisciplinary nature of SSM, showcasing its intersections with soft computing, behavior, and human capital management. These clusters represent not only distinct research areas but also integrated facets of a holistic approach to problem-solving. The amalgamation of computational acumen, behavioral insights, and skilled human resources enriches the SSM paradigm, making it adaptable to an array of challenges. This analysis not only deepens our understanding of the existing literature but also offers a roadmap for future research. By recognizing the symbiotic relationship between SSM and these key domains, scholars are encouraged to explore innovative applications, delve into the psychological nuances of decision-making, and invest in human capital development. This holistic perspective not only enhances the robustness of SSM applications but also contributes significantly to diverse fields, fostering a more informed, agile, and human-centric approach to problem-solving in the complex, interconnected world of today.

It appears that SSM's holistic approach, emphasizing systemic thinking, participatory engagement, and multidimensional problem-solving, positions it as a valuable tool in achieving sustainable development and SDGs. Its growing popularity in research literature reflects its effectiveness in addressing the multifaceted challenges of the modern world including such complex issues as human capital management. As organizations and policymakers seek innovative approaches to navigate the complexities of sustainable development, SSM stands out as a beacon, guiding them toward comprehensive, inclusive, and sustainable solutions.

In addition, this research demonstrated that SSM can be very helpful in sustainable human capital management practices which can enhance an organization's reputation as an employer of choice. Using the SSM approach, organizations can better assess their existing human capital management practices and pinpoint pathways for improvement. In today's competitive job market where attracting top talent is crucial for success, companies that prioritize sustainability are viewed more favorably by potential candidates seekingmeaningful employment experiences. A strong employer brand built on sustainable human capital practices can help attract high-caliber candidates who align with theorganization's values. Furthermore, enhancing sustainability in human capital management contributes to long-term organizational resilience. By investing in employee development and well-being, organizations can foster a culture of continuous learning and adaptability. Sustainable human capital management that employs SSM practices ensure that organizations have the right talent in place to navigate challenges and seize opportunities.

When it comes to the pathways for further research on SSM and its role in the modern complex world riddled by numerous challenges such as managing scarce natural and human resources, future reseach can focus on the wider bibliometric analysis encompassing such databases as Google Scholar or using the PlumX metrics that measure the social impact of every research. This approach, especially combined with the focus and scope of SSM might yield other interesting and relevant results that might be of special interest for relevant researhers, stakeholders, and policymakers.

**Funding:** This research was supported by the Russian Science Foundation grant No. 23-28-01768.

**Conflicts of Interest:** The authors declare no conflict of interest.